\documentstyle[aps,prl,twocolumn,epsf]{revtex}

\begin{document}

\twocolumn[\hsize\textwidth\columnwidth\hsize\csname @twocolumnfalse\endcsname

\draft

\title{End-Chain Spin Effects in Haldane Gap Materials}

\author{M. W. Meisel}
\address{Department of Physics, University of Florida, \mbox{P.O. Box 118440}, Gainesville,
 FL 32611-8440.}

\date{16 August 1998}

\maketitle

\begin {abstract}

This paper overviews the behavior of the end-chain spins of 
linear chain systems possessing a Haldane gap.  The physical properties of the end-chain 
spins are described by reviewing the results obtained primarily with materials known as NENP, 
Ni(C$_{2}$H$_{8}$N$_{2}$)$_{2}$NO$_{2}$(ClO$_{4}$), and NINAZ, 
Ni(C$_{3}$H$_{10}$N$_{2}$)$_{2}$N$_{3}$(ClO$_{4}$).

\end {abstract}

\pacs{PACS numbers: 76.30.-v, 75.10.Jm, 75.50 -y}

\vskip2pc]

\section*{Introduction}

Since the birth of quantum mechanics, low dimensional magnetic systems have been 
the subject of numerous theoretical and experimental investigations \cite{Bethe,Jongh,Steiner}.
In 1983, Haldane \cite{Haldane1} suggested that a one-dimensional Heisenberg 
antiferromagnet with integer spin possessed an energy gap between the nonmagnetic 
ground state and the first excited one.  In addition, this Haldane gap was not present 
in noninteger spin systems.  Haldane's humdinger prediction renewed interest in the field 
where most of the results may be explained by a spin Hamiltonian written as
\[
{\cal H} = J \sum_{i} \vec{S}_{i} \cdot \vec{S}_{i+1} + D \sum_{i} (S_{i}^{z})^{2} + 
E \sum_{i} [(S_{i}^{x})^{2}-(S_{i}^{y})^{2}]
\]
\begin{equation}
- \mu_{B} \vec{H} \cdot \sum_{i}  \bar{\bar{g}}_{i} \cdot \vec{S}_{i} +
J^{\prime} \sum_{j} \vec{S}_{j} \cdot \vec{S}_{j+1} \;\; ,
\end{equation}
where $J$ is the nearest-neighbor spin intrachain interaction, $D$ is the single-ion
anisotropy, 
$E$ is the orthorhombic anisotropy, $H$ is the externally applied magnetic field, and $J^{\prime}$ 
is the nearest-neighbor spin interchain interaction.

The purpose of this paper is not to provide a detailed review of theoretical and experimental 
work that lead to the identification of the Haldane phase.  Rather, this paper overviews the 
behavior of the end-chain spins of linear chain systems possessing a Haldane gap.  
Due to space limitations, this discussion cannot be comprehensive and will focus on aspects of 
our experimental work, since a few general reviews have been published 
\cite{AffleckR1,RenardR1,RenardR2,HalperinR1,ShibaR1,KatsumataR1}.  Finally, this paper will be 
organized in a quasi-chronological manner, so the reader may receive a sense of how the subject has evolved.

\begin{figure}[ht]
\epsfxsize=8.5cm \epsfbox{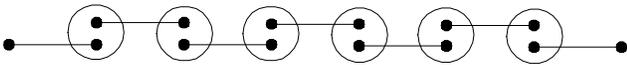}
\bigskip
\caption{A VBS model schematic representation of an $S = 1$ chain.}
\label{figure1}
\end{figure}

\section*{Prediction and Observation of End-Chain Spins in NENP}
The valence bond solid (VBS) model, introduced by Affleck, Kennedy, Lieb, and Tasaki \cite{Affleck1}, provides a physically intuitive picture of integer spin antiferromagnetic chains.  
For example as illustrated in Fig. 1, an $S = 1$ site has two $S = \frac{1}{2}$ entities, as 
represented by two points enclosed in a circle, that are coupled to neighboring sites by 
valence bonds represented by a solid line.  When compared to the resonating valence 
bonds \cite{Anderson2} of an $S = \frac{1}{2}$ chain, no degeneracy is available 
in forming the $S=1$ chain, so the singlet ground state is a solid.  Naturally at the ends 
of the chains, $S = \frac{1}{2}$ spins are predicted by the model.

Initial evidence of the presence of the end-chain spins was an enhanced impurity tail in 
the magnetic susceptibility of NENP, Ni(C$_{2}$H$_{8}$N$_{2}$)$_{2}$NO$_{2}$(ClO$_{4}$) 
\cite{Meyer,Renard1}, doped with Cu$^{2+}$ magnetic impurities \cite{RenardR1}.  Later, 
using X-band ESR on a similar sample, Hagiwara $et$ $al.$ \cite{Hagiwara1} observed hyperfine 
interactions that could be attributed to the interactions between $S = \frac{1}{2}$ variables 
and the Cu$^{2+}$ $S = \frac{1}{2}$ impurities.  Furthermore, the temperature dependence of 
the main ESR peak, $i.e.$ the peak attributed to the end-chains spins, was fit with an 
empirical formula whose form was later confirmed by Mitra, Halperin, and Affleck \cite{Mitra}.  
In order to break the chains while avoiding magnetic impurities, Glarum $et$ $al.$ \cite{Glarum} 
studied samples of NENP doped with Cd, Zn, and Hg.  These workers observed various asymmetric 
ESR peaks, depending upon the orientation of the magnetic field with respect to the crystal axes.
Recently, Ajiro $et$ $al.$ \cite{Ajiro} have reported a more extensive study of this doped system, 
and the asymmetries of the ESR signals were attributed to the staggered magnetization 
present in the sample.  

In our first experimental study \cite{Avenel1} of Haldane materials, we used single crystals 
of NENP and measured the magnetic susceptibility from 300 K down to below\linebreak
300 $\mu$K, Fig. 2.  
Our initial goal was to test the prediction \cite{Affleck3} that the $J^{\prime} / J$ value 
of NENP ($J^{\prime}/J \approx 10^{-4}$ and $J \approx 46$ K \cite{Renard1,Regnault}) was smaller 
than a critical value, so the material would remain disordered down to $T = 0$.  Our work revealed 
the presence of end-chain spins below 100 mK, and the corresponding signal established a roadblock for 
searching for long-range magnetic order.  Nevertheless, circumstantial evidence \cite{Avenel1} of some 
type of ordering is provided by a hysteresis in the signal while cooling-warming near 4 mK, Fig. 2.  
In an attempt to shift this transition to a higher temperature, a system with 
a larger $J$ value was needed, and NINAZ,  Ni(C$_{3}$H$_{10}$N$_{2}$)$_{2}$N$_{3}$(ClO$_{4}$) 
\cite{RenardR1,RenardR2,Granroth}, with $J \approx 125$ K \cite{Zheludev}, was chosen.

\begin{figure}[t]
\epsfxsize=8.5cm \epsfbox{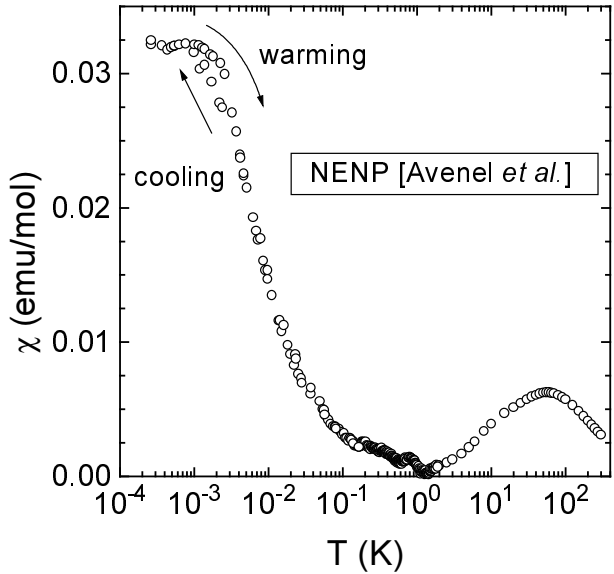}
\bigskip
\caption{The $\chi(T)$ of NENP over 6 decades of $T$ [19].  
Moving from high to low 
$T$, the data show a peak at $T \sim J$, an exponential decrease down to  
1 K, an increase below 100 mK, and hysteresis below 4 mK.}
\label{figure2}
\end{figure}

\section*{Evidence of End-Chain Spins in other Haldane Materials}
While we were studying NINAZ, other reports about end-chain spins were being made.  
The presence of $S=\frac{1}{2}$ end-chains spins on an $S=1$ chain 
was challenged by Ramirez $et$ $al.$ \cite{Ramirez} who 
performed specific heat measurements on pure and doped samples of Y$_{2}$BaNiO$_{5}$.  These 
results were interpreted to suggest that, for chains up to approximately 250 spins long, the 
end-chains were $S = 1$ entities existing in either in a singlet or triplet ground state, 
depending on whether the chain consisted of an even or odd number of spins.  Numerical 
work \cite{White1,Yamamoto} suggests that an even/odd chain length effect is not expected for 
chains longer than approximately 50 spins.  In fact, the data of Ramirez $et$ $al.$ have been 
described by $S=\frac{1}{2}$ end-chains experiencing a strong anisotropy \cite{Desmaris,Hallberg}.

It is important to stress that although subsequent experiments 
\cite{Kikuchi,Deguchi,Hirota,Fujiwara,Ito}
have been interpreted in terms of $S=\frac{1}{2}$ end-chains, these studies have not 
conclusively eliminated the $S = 1$ end-chain spin description asserted by Ramirez $et$ $al.$ 
\cite{Ramirez}.  
The reason for this ambiguity is that the ESR work was restricted to low frequencies where $S=1$ end-chain 
spins could not be measured, if they were present.  Furthermore, the temperature dependence of the 
central peak of the ESR signal in doped samples of NENP \cite{Hagiwara1,Mitra} and of 
TMNIN \cite{Deguchi,Ito}, which is  
(CH$_{3}$)$_{4}$N[Ni(NO$_{2}$)$_{3}$] \cite{Chou}, have been described by the model of 
Mitra, Halperin and Affleck \cite{Mitra}.  However, since the lineshapes are
severely distorted in 
doped samples, the 
linewidths, and therefore the interactions between the end-chain spins on the same chain, could not
be observed.  Although nonmagnetic dopants are used to increase the end-chain spin concentration, 
doped samples have their limitations.  For example in NENP doped beyond $\sim 0.5$\%, the dopant 
no longer breaks chains by direct substitution \cite{Fujiwara}.  Even in a material with a 
nonmagnetic isomorph \cite{Kikuchi}, 
doping beyond a certain level is nonlinear with respect to the number of observed end-chain spins.  
Furthermore, it is important to stress that dopants cause changes in the magnetic environment and may 
shift, split, and/or broaden the ESR spectra.  
The material known as NINAZ allows us to avoid the doping difficulties and 
is well-suited to study the $S=\frac{1}{2}$ versus $S=1$ issue.  In addition, 
the magnetic properties of NINAZ may be related to the interactions between the end-chain 
spins on the shortest chains and between the magnetic excitations on the chains and the end-chain spins.

Finally, it is noteworthy that end-chain spin effects are not restricted to $S=1$ systems.  The VBS model 
\cite{Affleck1} may be generalized, and $S=1$ end-chain spins are expected for $S=2$ Haldane 
gap materials \cite{Nishiyama,Schollwock,Yamamoto2}.  Experimental evidence of these $S=1$ end-chain spins has been obtained in the $S=2$ 
linear-chain materials of MnCl$_{3}$(bipy) \cite{Granroth2} and 
CsCrCl$_{3}$ doped with Mg \cite{Yamazaki}.

\section*{End-Chain Spins in NINAZ}

The material NINAZ circumvents difficulties associated with doping because it shatters while 
passing through a structural transition at $\sim 255$ K, thereby producing end-chain spins 
without doping \cite{RenardR2,Granroth,Chou}.  Since the nascent crystals shattered upon cooling but
remained oriented, these specimens are referred to as \textit{polycrystalline} samples.  To increase the 
number of end-chain spins, two grinding techniques were used.  Initial grindings 
used a pestle and mortar to produce a sample referred to as \textit{powder}.  Subsequent grindings, using 
a standard ball mill, yielded a sample referred to as \textit{ultrafine powder}.

Electron spin resonance at 9, 94, and 190 GHz, and magnetization studies on all of the samples 
confirmed that the end-chain spins are $S = \frac{1}{2}$ and show no evidence for $S=1$ 
end-chains \cite{Granroth}.  
More specifically, the magnetization data as a function of magnetic field up to 5 T 
while at a temperature of 2 K may be fit using the sum of two Brillouin functions, one for 
$S=\frac{1}{2}$ and the other for $S=1$.  The fits of the data demonstrate that the end-chain 
spins are predominately $S=\frac{1}{2}$ and that only trace amounts of $S=1$, consistent with 
the presence of some uncoupled Ni$^{2+}$, exist in any of the samples.  Furthermore, the high 
frequency ESR studies did not reveal the presence of any $S = 1$ or higher spin contributions.  
Consequently, we conclude that all the end-chain spins are $S=\frac{1}{2}$.

For our 9.25 GHz ESR work, the measured derivative spectra of all the samples 
were integrated at a variety of temperatures \cite{Granroth}.  
The data were fit with the expressions generated by Mitra, Halperin, and Affleck \cite{Mitra} and 
yield values for two parameters, namely the average chain length ($L_{0}$) and the minimum 
chain length ($L_{min}$).  The values of $L_{0}$ for the different samples are in reasonable 
agreement with the ones independently obtained from fits of the ESR and magnetization data.  
For the analysis of the ESR data, $L_{min} = 60$ $\pm$ $20$ sites was used.

Typical 9.25 GHz ESR lines acquired at 4 K are shown in Fig. 3, where the full width half maxima 
(FWHM) for each sample is approximately 8 mT \cite{Granroth}.  
For the purpose of comparison, the ESR spectrum for 
a 0.5\% Hg doped polycrystalline specimen is shown as an inset to Fig. 3.  These results indicate that 
the signals are dependent on 
\begin{figure}[ht]
\epsfxsize=8.5cm \epsfbox{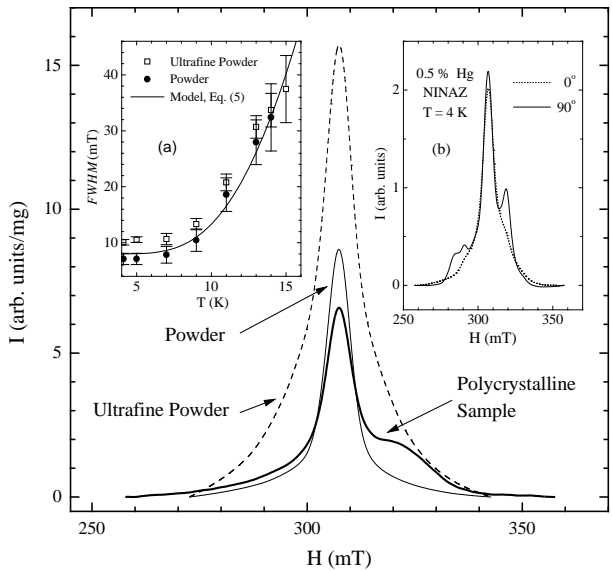}
\bigskip
\caption{Typical 9.25 GHz line shapes of NINAZ at 4 K [22].  
Inset (a) shows the temperature dependence of the FWHM, 
and the line is a fit using 
Eq. (3).  Inset (b) shows the signal from a 0.5\% Hg doped sample.}
\label{figure3}
\end{figure}
orientation and demonstrate why linewidth measurements of doped 
materials have not been reported in detail ($i.e.$ doping broadens the main line and adds 
extrinsic ones).  The temperature dependencies  of the linewidths are shown as another inset to 
Fig. 3.  For $T \leq 7$ K, the FWHM is temperature independent, and this limit is governed by 
the interactions between the $S=\frac{1}{2}$ end-chains on the shortest chains.  Following 
the model of Mitra, Halperin, and Affleck \cite{Mitra}, this interaction may be roughly 
estimated as
\begin{equation}
\epsilon=\Delta \exp (-L_{min}/\xi),
\end{equation}
where $\Delta (\approx 42$ K \cite{Zheludev}) is the Haldane gap and $\xi (\approx 6$ \cite{White1}) 
is the
correlation length.  When $L_{min} = 50$, then the resultant energy is
$\epsilon \approx$ 10 mK $\approx$ 8 mT.  Although we cannot completely eliminate the 
possibility that the temperature independent FWHM value is a consequence of dipole-dipole 
broadening, the estimates for $L_{min}$ are consistent with fits of the temperature dependence 
of the main peak and with numerical work \cite{White1,Yamamoto}.

When magnetic excitations are present on the chains, there are two mechanisms \cite{Mitra} by which
these bosons may influence the linewidth.
Firstly, bosons changing energy levels (via interactions with the end-chain spins)
could affect the linewidth, where this change in
energy is quantized in units of
\begin{equation}
\delta E \approx \frac{(\hbar \pi c)^{2}}{2 \Delta L^{2}},
\end{equation}
where $c$ ($= 2.55 \times 10^{4}$ m/s \cite{Zheludev}) is the speed of the spin wave 
and $L$ is the length of the chain.  
Using the independently measured \cite{Zheludev} values of $c$ and $\Delta$ and taking $L = L_{0}$, 
$\delta E \approx 5$ mT for the powder and
$\approx 12$ mT for the ultrafine powder.  These values of $\delta E$ are about the size of the FWHM, and
consequently, these interactions contribute to the linewidth once a chain acquires a boson.  The temperature
dependence of the linewidth near the central peak has been derived by Mitra, Halperin, and Affleck
\cite{Mitra}, so the FWHM may be modeled
by
\begin{equation}
\mbox {FWHM} = \epsilon + \Lambda \: T \exp(-\Delta/k_{B}T),
\end{equation}
where $\Lambda$ is a parameter which is beyond the scope of the present model \cite{Mitra}.
The temperature dependence
of the FWHM is reasonably reproduced when $\epsilon = 8$ mT and $\Lambda = 35$ mT/K, as shown in Fig.\ 3a.
In other words, the
major contribution to the FWHM, at $T \leq 7$ K, comes from the end-chain spins on the shortest chains
interacting with each other and, at $T \geq 7$ K, arises from the interactions between the magnetic
excitations on the chains and the end-chain spins.
Finally, the other possible broadening mechanism arises from a small change in the energy of a boson that
experiences a phase shift when interacting with an end-chain spin.
This effect is significantly weaker than the exchange of energy $\delta E$ and,
consequently, is not detectable in our measurements.

\section*{Acknowledgments}
I am deeply indebted to my numerous coworkers \cite{Avenel1,Granroth,Zheludev,Chou,Granroth2} 
who have made our experiments possible, especially G. E. Granroth, S. E. Nagler, and 
D. R. Talham.  Early aspects of this work were supported, in part, by the National Science 
Foundation (NSF DMR-9200671).  This manuscript is supported, in part, by a collaboration (NSF INT-9722935) 
with A. Feher (P. J. \v{S}af\'{a}rik University, Ko\v{s}ice, Slovakia), and I am grateful for his 
generous hospitality.

\end{document}